\documentclass[twocolumn,prl,superscriptaddress,amsmath,amssymb,10pt]{revtex4}

\usepackage{graphicx}
\usepackage{dcolumn}
\usepackage{stmaryrd}

\begin{document}

\title{{\normalsize Emergence of Chaotic Itinerancy in Simple Ecological Systems}}

\author{Pan-Jun Kim}
\affiliation{Department of Physics, Korea Advanced Institute of Science and Technology, Daejeon, Korea}
\author{Tae-Wook Ko}
\altaffiliation{Presently at the Department of Mathematics, University of Pittsburgh, Pittsburgh, PA 15260, USA.}
\affiliation{National Creative Research Initiative Center for Neuro-dynamics and Department of Physics, Korea University, Seoul, Korea}
\author{Hawoong Jeong}
\affiliation{Department of Physics, Korea Advanced Institute of Science and Technology, Daejeon, Korea}
\author{Kyoung J. Lee}
\affiliation{National Creative Research Initiative Center for Neuro-dynamics and Department of Physics, Korea University, Seoul, Korea}
\author{Seung Kee Han} \email{skhan@chungbuk.ac.kr}
\affiliation{Department of Physics, Chungbuk National University, Cheongju, Chungbuk, Korea}

\date{\today}

\begin{abstract}
Chaotic itinerancy is a universal dynamical concept that describes itinerant motion among many different ordered states through chaotic transition in dynamical systems. Unlike the expectation of the prevalence of chaotic itinerancy in high-dimensional systems, we identify chaotic itinerant behavior from a relatively simple ecological system, which consists only of two coupled consumer-resource pairs. The system exhibits chaotic bursting activity, in which the explosion and the shrinkage of the population alternate indefinitely, while the explosion of one pair co-occurs with the shrinkage of the other pair. We analyze successfully the bursting activity in the framework of chaotic itinerancy, and find that large duration times of bursts tend to cluster in time, allowing the effective burst prognosis. We also investigate the control schemes on the bursting activity, and demonstrate that invoking the competitive rise of the consumer in one pair can even elongate the burst of the other pair rather than shorten it.
\end{abstract}

\maketitle
Since attractors determine the long-term behavior of dynamical systems, the concept of attractors is central to the analysis of many natural and artificial systems \cite{Strogatz}. In general, the phase space of a nonlinear dynamical system is partitioned into various basins of attraction from which states evolve towards the respective attractors. These stable attractors can lose their stability with a change of the system condition such that the basin of attraction of each attractor becomes connected to each other through unstable manifolds. Hence, a dynamical state which sequentially traces out all of the destabilized {\it attractor ruins} emerges. This is referred to as a chaotic itinerant state \cite{Kaneko}. The notion of chaotic itinerancy has received considerable attention in studying the adaptability of complex systems, especially in relation to brain information processing \cite{Kaneko,Tsuda}.

To embed a chaotic itinerant state, a system is expected to have a high degree of complexity; therefore, models of chaotic itinerancy are mostly built on high-dimensional phase space \cite{high}. Albeit relatively low-dimensional systems, two coupled Morris-Lecar neural oscillators were found to exhibit chaotic itinerancy \cite{Han}, the result seems to be limited to a rather special case obtained by using sophisticated forms of model equations in neurobiological systems. In the present work, we report that low-dimensional chaotic itinerancy exists and arises naturally in simple ecological systems, of which consumer-resource dynamics has broad relevance in metabolic, immune, social, and economical systems. The wide variety of related disciplines aside, the mathematical simplicity of our low-dimensional system renders the global organization of a chaotic itinerant state tractable with a detailed illustration.

At the outset, we suggest the equations of two consumer-resource pairs coupled via resource sharing \cite{origin}:
\begin{eqnarray}\label{rdc}
\frac{dC_{1(2)}}{dt} &=& aC_{1(2)}\left[\frac{R_{1(2)}}{\kappa + R_{1(2)}}+\frac{DR_{2(1)}}{\kappa + R_{2(1)}}\right] -bC_{1(2)} \,,\nonumber\\
\frac{dR_{1(2)}}{dt} &=& R_{1(2)} - R^{2}_{1(2)}-\frac{[C_{1(2)}+DC_{2(1)}]R_{1(2)}}{\kappa + R_{1(2)}} \,,
\end{eqnarray}
where $C_{1(2)}$ and $R_{1(2)}$ represent the population of consumer 1 (2) and resource 1 (2), respectively. $a$ and $b$ denote the growth and decay rates of the population of consumers. $\kappa$ concerns the satiability level of the consumers taking resources. For simplicity in our analysis, we do not distinguish the parameter sets of the two consumer-resource pairs. In this equation, $R_{1(2)}$ is taken by $C_{1(2)}$ primarily, as well as by $C_{2(1)}$ with a relative small uptake rate $D$, which ranges from $0$ to $1$.

\begin{figure}[t]
\begin{center}
\includegraphics[width=0.475\textwidth]{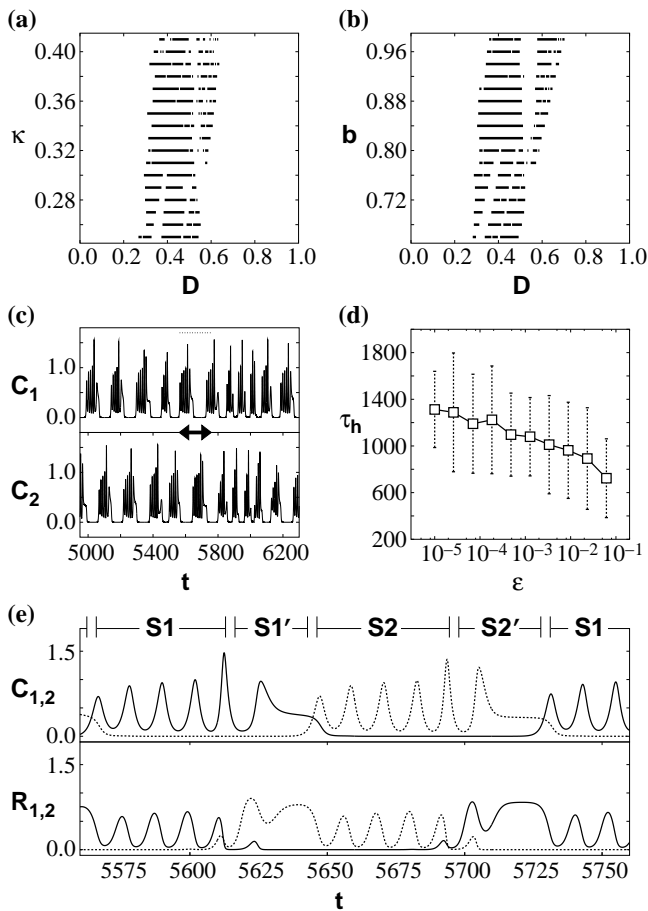}
\caption{(a) and (b): parameter regime of irregular bursting activities in Eq.~(\ref{rdc}) with $a=2$, when (a) $b=0.82$ or (b) $\kappa=0.33$. (c) Time series of $C_{1}$ and $C_{2}$ in bursting activity. The arrowed time interval is magnified in (e). (d) Half-life time $\tau_{h}$ of burst reproducibility with initial condition difference $\varepsilon$. (e) Magnification of the arrowed time interval in (c). Solid lines denote $C_{1}$ and $R_{1}$, and dotted lines $C_{2}$ and $R_{2}$.
}
\label{fig1}
\end{center}
\end{figure}

When $D$ is equal to zero, Eq.~(\ref{rdc}) splits into two Holling type-II forms of Lotka-Volterra equations \cite{Holling},
and the populations of each consumer-resource pair can exhibit a limit cycle oscillation. If $D$ takes a nonzero value close to $0$ or $1$, synchronous limit cycle oscillation between the two consumer-resource pairs arises. Complicated dynamics develop at intermediate range of $D$, where we can observe irregular bursting activities as in Figs.~\ref{fig1}(a)--\ref{fig1}(c). Time trajectories of $C_{1}$ and $C_{2}$ in Fig.~\ref{fig1}(c) show the bursting behaviors, and those of $R_{1}$ and $R_{2}$ show the similar patterns as well. $C_{1}$ and $C_{2}$, or $R_{1}$ and $R_{2}$, fire bursts in an antiphase-synchronized way, such that the explosion of $C_{1}$ or $R_{1}$ co-occurs with the shrinkage of $C_{2}$ or $R_{2}$, and vice versa. It is worth noting that other equations with similar systems to Eq.~(\ref{rdc}) also support the existence of such antiphase-synchronized irregular bursts \cite{Vand}. For numerical simulations, we use the parameters $a=2$, $b=0.82$, $\kappa=0.33$, and $D=0.57$ unless specified.

To check whether the apparent irregularity of the bursts implies their initial condition sensitiveness, we evaluate
\begin{equation}\label{Ls}
M_{1(2)}(t)=\frac{1}{t}\int_{0}^{t}H[C_{1(2)}(t')]H[C'_{1(2)}(t')]dt'\,,
\end{equation}
where $H(X)=1$ if $X$ is in the burst mode, otherwise $H(X)=-1$. The antiphase synchronization of the bursts enables us to define the burst mode unambiguously, such that once $C_{1(2)}$ exceeds $C_{2(1)}$, $C_{1(2)}$ enters the burst mode, and $C_{2(1)}$ enters the shrinkage mode. $C'_{1(2)}(t)$ is calculated in the same way as $C_{1(2)}(t)$, but is initially perturbed from $C_{1(2)}(t)$ with $\varepsilon =|C'_{1(2)}(0)-C_{1(2)}(0)|/C_{1(2)}(0)\ll 1$. Therefore, Eq.~(\ref{Ls}) gives the similarity between the bursting times of $C_{1(2)}(t)$ and those of $C'_{1(2)}(t)$ with slightly different initial conditions. If the bursting times of $C_{1(2)}(t)$ and $C'_{1(2)}(t)$ are in complete agreement, $M_{1(2)}(t)=1$, whereas with no correlation between them, $M_{1(2)}(t)=0$. In the following, we drop the subscript of $M_{1(2)}(t)$ due to the statistical equivalence of $C_{1}(t)$ and $C_{2}(t)$. One can employ $M(t)$ for determining the necessary time for the discrepancy to be significant between the bursting times of $C_{1(2)}(t)$ and of $C'_{1(2)}(t)$. It is observed that $M(t)$ evolves rapidly from $1$ to $0$ in the irregular bursting regime; thus, the half-life time $\tau_{h}$ of $M(t)$ can serve as the characteristic time scale of the discrepancy growth. Figure~\ref{fig1}(d) shows that $\tau_{h}$ scales logarithmically to $\varepsilon$, and using $R_{1(2)}(t)$ and $R'_{1(2)}(t)$ instead of $C_{1(2)}(t)$ and $C'_{1(2)}(t)$ in Eq.~(\ref{Ls}) does not alter the current result. This logarithmic scaling reveals that the bursting is sensitive to the initial conditions, i.e., behaves chaotically \cite{Liap}.

To address such antiphase-synchronized chaotic bursts in detail, we divide a period of bursts into four stages - $\rm S1$, $\rm S1'$, $\rm S2$, and $\rm S2'$, as in Fig.~\ref{fig1}(e). In stage $\rm S1$, $C_{1}$ and $R_{1}$ dominate $C_{2}$ and $R_{2}$, while $C_{1}$, which is supported primarily by $R_{1}$, depresses severely the growth of $R_{2}$, and thereby of $C_{2}$. Nonetheless, $R_{2}$ increases gradually in the negligible presence of $C_{2}$, and $R_{1}$ comes to decline with overpopulated $C_{1}$ which takes both $R_{1}$ and $R_{2}$. In stage $\rm S1'$, the resultant shrinkage of $R_{1}$ ensures that $C_{1}$ depends mostly on $R_{2}$ for survival. Meanwhile, $R_{2}$ can boost the increase of $C_{2}$, which then suppresses both $R_{2}$ and $R_{1}$, thereby leading to the drastic decay of $C_{1}$ in stage $\rm S2$ \cite{shrink}. The dominance of $C_{2}$ and $R_{2}$ in stage $\rm S2$ is totally symmetric to that of $C_{1}$ and $R_{1}$ in stage $\rm S1$. Accordingly, stage $\rm S2'$ analogous to stage $\rm S1'$ follows, and leads to stage $\rm S1$ for the next period.

\begin{figure}[t]
\begin{center}
\includegraphics[width=0.475\textwidth]{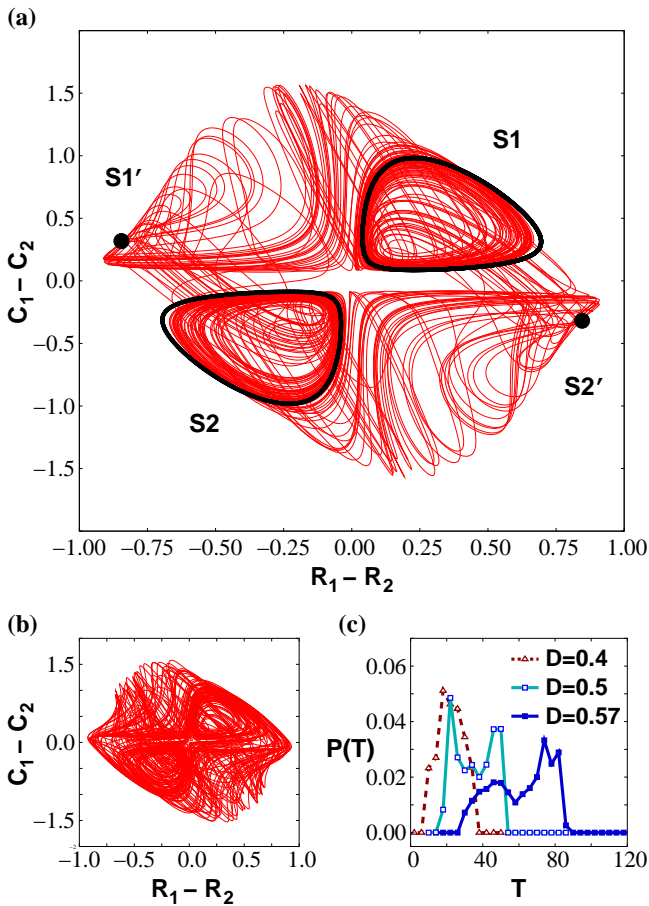}
\caption{(a) Projection of time trajectory onto $(R_{1}-R_{2}, C_{1}-C_{2})$ for $\Delta t=3000$, $D=0.57$ (thin line, red in color). Overlapped is invariant solution at each stage (black). (b) Depicted by the similar way with (a), when $D=0.4$. (c) Distribution of burst duration $T$ of $C_{1}$ with stages $\rm S1$ and $\rm S1'$, or of $C_{2}$ with stages $\rm S2$ and $\rm S2'$, for different values of $D$ ($C_{1}$ and $C_{2}$ show the identical distribution). Similar results are also observed with each of the stages.
}
\label{fig2}
\end{center}
\end{figure}

Each stage occupies finite time-span, forming a quasi-stable dynamical state. The alternating dominance of each species along the stages may be equivalent to the switching events among the sets of attractor ruins. In order to elucidate the underlying attractor ruin for a given stage, we consider an invariant subspace of $(C_{1},R_{1},C_{2},R_{2})$ which contains only the species governing the stage [see Fig.~\ref{fig1}(e)]: at stage $\rm S1$, $(C_{1},R_{1},0,0)$; at stage $\rm S1'$, $(C_{1},0,0,R_{2})$; at stage $\rm S2$, $(0,0,C_{2},R_{2})$; at stage $\rm S2'$, $(0,R_{1},C_{2},0)$. The populations confined within each invariant subspace approach their own asymptotic solution. For instance, the limit cycle oscillation of $C_{1}$ and $R_{1}$ characterizes the asymptotic solution in the invariant subspace of stage $\rm S1$ and thus underlies the bursting activity at this stage. Figure~\ref{fig2}(a) shows an actual time trajectory of the populations in phase space, which also embeds the asymptotic solutions in the invariant subspaces. Near an invariant asymptotic solution, the trajectory remains there for a long time, but finally escapes towards another invariant solution at the next stage.

This escaping event is due to an existence of unstable manifolds outward from an invariant subspace. Along the transverse direction of the invariant subspace, we then perform a linear stability analysis to find the unstable manifolds:
\begin{eqnarray}\label{mani}
{\rm S1:}\,\,\,\,
\frac{d\delta R_{2}}{dt} &\cong& \left(1-D\frac{C_{1}}{\kappa}\right) \delta R_{2} \,,\nonumber\\
{\rm S1':}\,\,\,\,
\frac{d\delta C_{2}}{dt} &\cong& \left[{\rm min}\left(\frac{b}{D}\,,\frac{a}{1+\kappa}\right)-b\right]\delta  C_{2} \,,
\end{eqnarray}
where $C_{1}$ of stage $\rm S1$ is evaluated in the absence of $C_{2}$ and $R_{2}$. Stages $\rm S2$ and $\rm S2'$ take the formula obtained simply by interchanging $C_{1}$ and $C_{2}$, $R_{1}$ and $R_{2}$ of stages $\rm S1$ and $\rm S1'$ in Eq.~(\ref{mani}). The formula of unstable manifolds reveals which species causes the instability of a given stage; the instability of stages $\rm S1$ and $\rm S1'$ is invoked by the increase of $R_{2}$ and $C_{2}$, as described above in the ecological argument. From Eq.~(\ref{mani}), we recognize that increase of $D$ tends to reduce the coefficients in the right-hand sides, i.e., to decrease the escape rates along the unstable manifolds. The resultant relaxation of the switching events among attractor ruins is identified by comparing Figs.~\ref{fig2}(a) and \ref{fig2}(b), where consistently the trajectory between the invariant solutions looks less intermingled with increased $D$. The most evident effect of elongated residence in attractor ruins appears in Fig.~\ref{fig2}(c): the distribution of burst duration $T$ shifts to large value as $D$ increases. Moreover, the larger $D$ is, the higher the right peak of the bimodal duration distribution is, relatively to the left peak. We conclude that in the chaotic bursting regime, the enhanced coupling strength induces either of the consumer-resource pairs to dominate the other for a long time by an elongated bursting activity.

\begin{figure}[t]
\begin{center}
\includegraphics[width=0.475\textwidth]{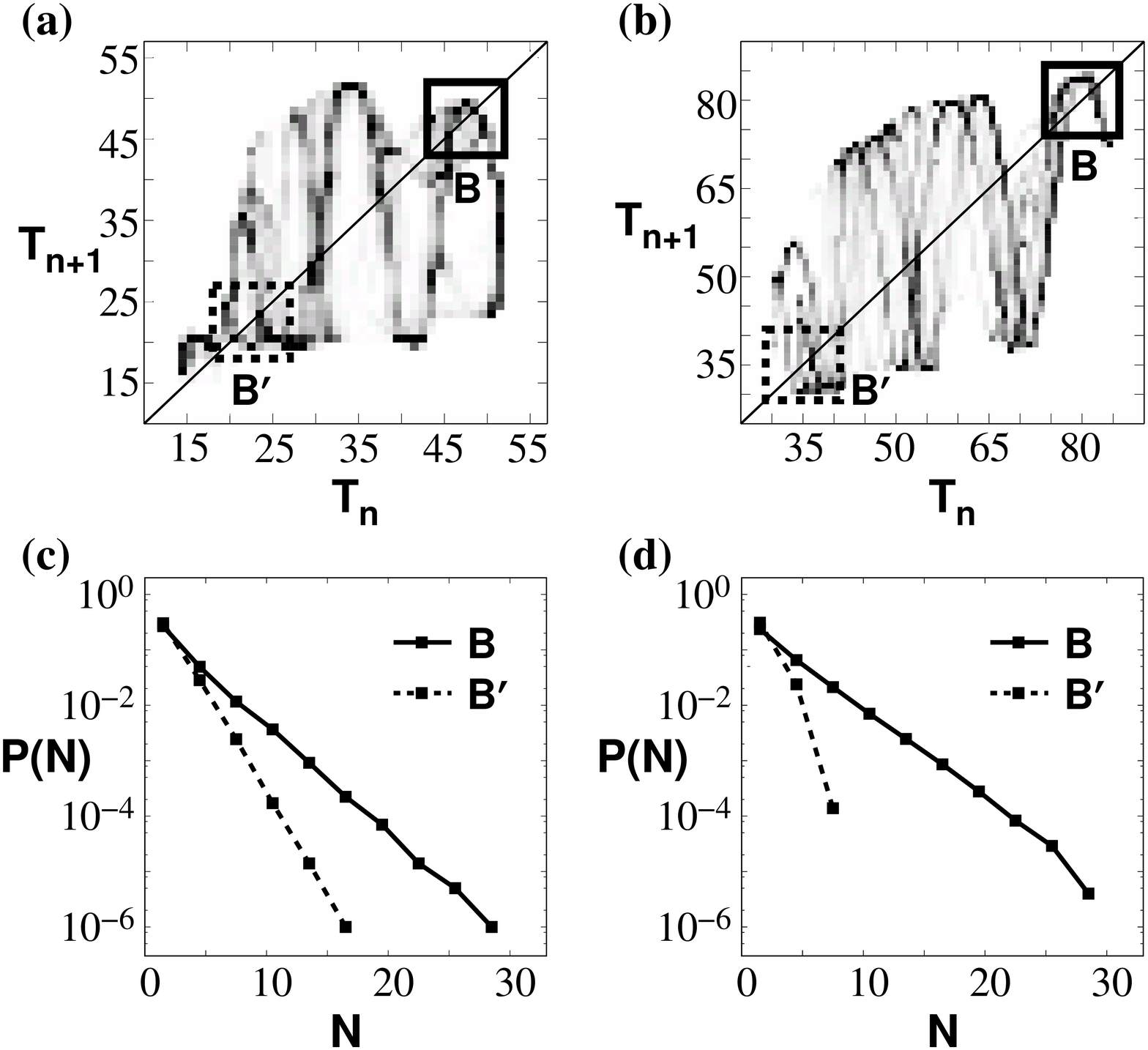}
\caption{(a) and (b): return time map for burst initiations when (a) $D=0.5$ or (b) $D=0.57$. (c) and (d): distribution of residual times $N$ within boxed area in (a) [(c)] or in (b) [(d)].
}
\label{fig3}
\end{center}
\end{figure}

To investigate the bimodality appearing in Fig.~\ref{fig2}(c), we plot a return time map for burst initiations of consumers by considering the terms between the initiation of stage $\rm S1$ and that of stage $\rm S2$, and between the initiation of stage $\rm S2$ and that of stage $\rm S1$, and so on. Figures~\ref{fig3}(a) and \ref{fig3}(b) display nearly one-dimensional curves of such return time maps, where stepwise jumps between the upper and lower extremes are responsible for the bimodality in Fig.~\ref{fig2}(c). The relative expansion of upper extremes in Fig.~\ref{fig3}(b) induces the right side of the duration distribution in Fig.~\ref{fig2}(c) to be highly peaked.

Referring to the return time maps, we find that mapping trajectories are frequently trapped in boxed area $\rm B$ in Figs.~\ref{fig3}(a) and \ref{fig3}(b). The distribution of residual times $N$ (total number of iteration) within boxed area $\rm B$ is indeed more right-skewed than those of any other areas (e.g., $\rm B'$) with the same size [Figs.~\ref{fig3}(c) and \ref{fig3}(d)]. The distribution within area $\rm B$ follows exponential fit $P(N)\propto p^{N}$ with $p=0.64$ for $D=0.5$ and $p=0.68$ for $D=0.57$, and shows significantly larger $p$ than the surrogated data ($p=0.33$ for $D=0.5$, $p=0.28$ for $D=0.57$) with the same distribution of burst duration time. This indicates the residual dynamics within area $\rm B$ follows a poissonian process, but with considerable survivability $p$ per iteration. Since area $\rm B$ corresponds to relatively long durations of bursts, large duration times of bursts tend to cluster in time, as partially observed in Fig.~\ref{fig1}(c). In this regard, the known information about the past duration could be beneficial to improve the burst prognosis efficiently.

\begin{figure}[t]
\begin{center}
\includegraphics[width=0.475\textwidth]{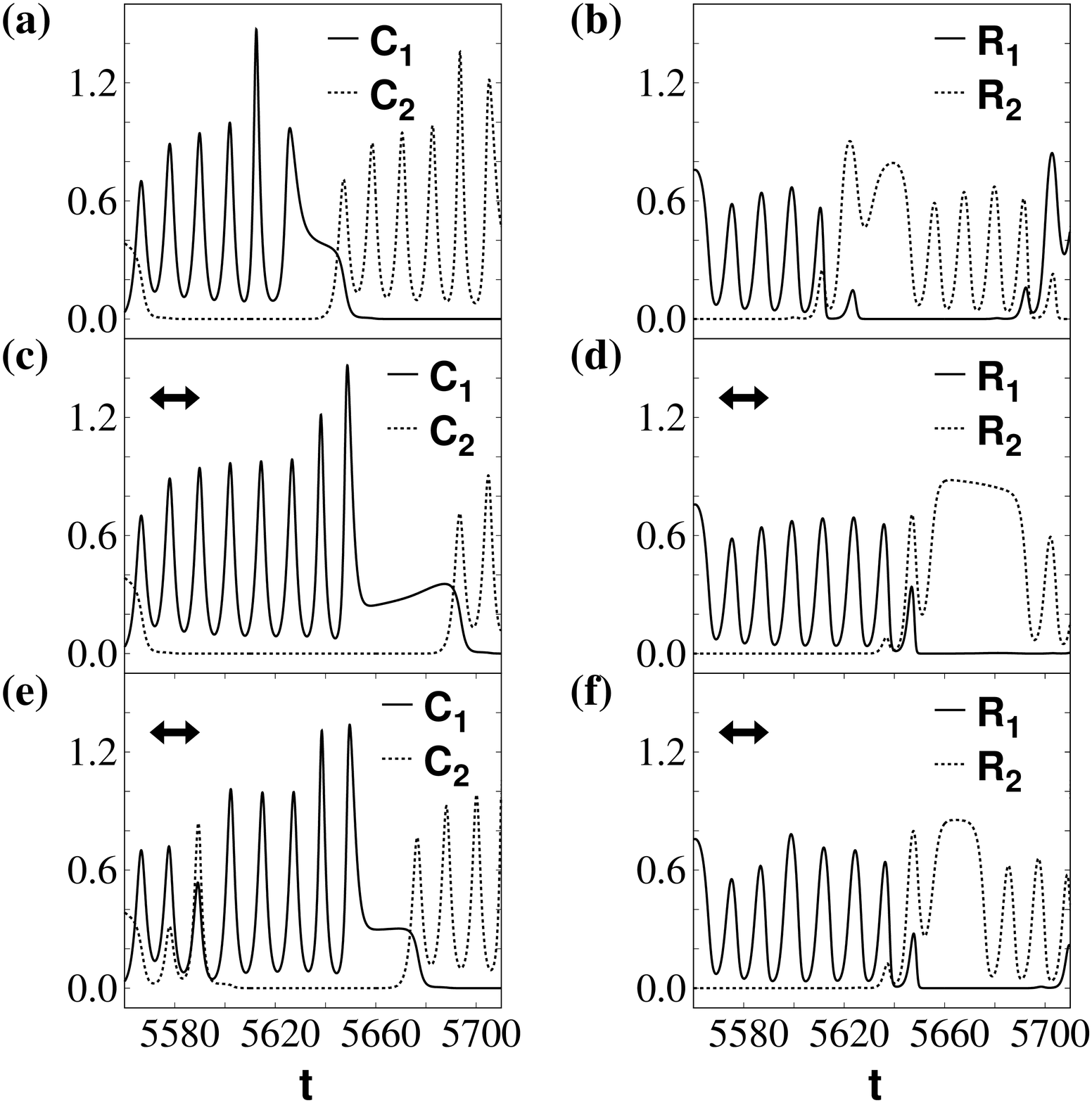}
\caption{(a) and (b): unperturbed time-series of populations. (c)--(f): initially the same as (a) and (b), but subjected to the perturbation during the arrowed time interval by reducing $R_{2}-R^{2}_{2}$ in half [(c), (d)] or doubling $aC_{2}$ [(e), (f)], given the terms in Eq.~(\ref{rdc}).
}
\label{fig4}
\end{center}
\end{figure}

An important outcome of the stability analysis on attractor ruins is the application to a control scheme on burst duration. Since a rise of $R_{2(1)}$ destabilizes the dominance of $C_{1(2)}$ and $R_{1(2)}$, manually repressing the growth of $R_{2(1)}$ might elongate the bursts of $C_{1(2)}$ and $R_{1(2)}$. As shown in Figs.~\ref{fig4}(c) and \ref{fig4}(d), the repressed growth of $R_{2}$ prevents the overpopulation of $C_{1}$ for a while and thus delays the shrinkage of $R_{1}$ as well as of $C_{1}$. It should be noticed that the effect of delayed rise of $C_{2}$ herein is not so essential to elongating the burst duration, despite the resource competition between $C_{2}$ and $C_{1}$. Counterintuitively, even promoting the growth of $C_{2}$ could be helpful to the burst duration, if resulting in the depression of $R_{2}$. Figures~\ref{fig4}(e) and \ref{fig4}(f) indeed illustrate the possibility that a sufficiently large perturbation to increase $C_{2}$ withdraws transiently $R_{1}$ and $C_{1}$, but also delays the growth of $R_{2}$ thereby elongating the bursting activity \cite{sa}.

In summary, we investigated a simple dynamical system, which consists only of two consumer-resource pairs but exhibits chaotic itinerancy naturally.
The mathematical simplicity of the system gives rise to a clear view of the organization of a chaotic itinerant state, where each consumer-resource relationship underlies its corresponding attractor ruin as a dynamical `building block'. Such concept of building blocks could be generically utilized when one designs other systems exhibiting chaotic itinerancy.
In addition, analysis on a chaotic itinerant state was found to be applicable to the prognosis and control of the dynamical system,
in the rather counterintuitive way.
Beyond the suggested ecological system, any dynamical system which shows antiphase-synchronized chaotic bursts might be analyzable in the framework of chaotic itinerancy via our methodology.
We expect that host-parasitoid systems with whiteflies and their parasitic wasps could be employed for experimental validation of our results, since parasitic wasps (consumers) are known to have overlapped hosts (resources) in a manner similar to the present model \cite{expr}.

The authors thank D. E. Postnov, Dong-Uk Hwang, Hwang-Yong Kim for useful discussion.
This work was supported by the Korean Systems Biology Research Project (M10309020000-03B5002-00000) (P.-J.K. and H.J.),
the Creative Research Initiatives of the Korean Ministry of Science and Technology (T.-W.K. and K.J.L.),
the SBD-NCRC program (R15-2004-033-07001-0), and the BK21 program at Chungbuk National University (S.K.H.).


\begin{thebibliography}{99}
\bibitem{origin}
 Equation~(\ref{rdc}) is obtained by rescaling of the following equation:
 \begin{eqnarray*}
 \frac{dC_{1(2)}}{dt} &=& aC_{1(2)}\left[\frac{R_{1(2)}}{\kappa + \mu R_{1(2)}}+\frac{DR_{2(1)}}{\kappa + \mu R_{2(1)}}\right] -bC_{1(2)} \,,\\
 \frac{dR_{1(2)}}{dt} &=& \sigma R_{1(2)} - \gamma R^{2}_{1(2)}-\lambda\frac{[C_{1(2)}+DC_{2(1)}]R_{1(2)}}{\kappa + \mu R_{1(2)}} \,,
 \end{eqnarray*}
 where $C_{1(2)}$ and $R_{1(2)}$ represent the population of consumer 1 (2) and resource 1 (2), respectively.
 The parameters include $a$, $\kappa$, $\mu$, $D$, $b$, $\sigma$, $\gamma$, $\lambda$, of which $\mu$, $\sigma$, $\gamma$, $\lambda$ can be safely removed to give Eq.~(\ref{rdc}) without loss of generality.
\bibitem{Liap}
 Because the conventional Lyapunov exponent does not resolve the time scales involved in the present dynamics, we can hardly capture whether the chaotic behavior emerges at slow bursting scales or at fast oscillating scales. To clarify the chaotic activity in slow time scales, we instead utilize Eq.~(\ref{Ls}), which gives eventually the divergence rate of nearby initial conditions at bursting time scales. From Fig. \ref{fig1}(d), one can obtain ${\rm constant} \cong \varepsilon e^{q\tau_{h}}$ where $q$ is expected to be the rough estimation of the maximum Lyapunov exponent $\lambda_{\rm max}$, as proven by $q=0.016$ and $\lambda_{\rm max}=0.023$.
\bibitem{shrink}
 If $D$ is small enough that $R_{2}$ alone cannot maintain $C_{1}$ as nonzero, $C_{1}$ becomes already decayed at stage $\rm S1'$. Otherwise, with sufficiently large $D$ at stage $\rm S1'$, $C_{1}$ can show either of stationary or oscillatory behavior depending on the parameters.
\bibitem{sa} 
 For Figs. \ref{fig4}(c)--\ref{fig4}(f), let $t_{p}$ denote the initial time of perturbation, $\Delta t_{p}$ the duration time of perturbation, and $f_{p}$ the relative decrease of $R_{2}-R^{2}_{2}$ [Figs. \ref{fig4}(c) and \ref{fig4}(d)] or increase of $aC_{2}$ [Figs. \ref{fig4}(e) and \ref{fig4}(f)] with Eq.~(\ref{rdc}). We change each of $t_{p}$, $\Delta t_{p}$, and $f_{p}$ while the others remain fixed as in Figs. \ref{fig4}(c)--\ref{fig4}(f), and observe the burst being elongated for Figs. \ref{fig4}(e) and \ref{fig4}(f) under the limitations $t_{p} \le 5573$, $4.5 \le \Delta t_{p}$, $0.8 \le f_{p}$. Either decreasing $t_{p}$, increasing $\Delta t_{p}$, or increasing $f_{p}$ tends to elongate the burst durations throughout Figs. \ref{fig4}(c)--\ref{fig4}(f).
\end{thebibliography}
\end{document}